\documentclass[twocolumn,aps,floats,preprintnumbers]{revtex4}
\usepackage{epsfig}
\usepackage{amsmath}
\usepackage{amsfonts}
\usepackage{amssymb}

\begin{document}

\preprint{MZ-TH/08-2}
\preprint{January 4, 2008}

\title{QCD Calculations of Decays of Heavy Flavor Hadrons}

\author{Matthias Neubert\footnote{On leave from Laboratory for Elementary-Particle Physics, Cornell University, Ithaca, NY 14853, U.S.A.}}

\affiliation{Institut f\"ur Physik (THEP), Johannes Gutenberg-Universit\"at\\ 
D--55099 Mainz, Germany}

\begin{abstract}
Precision tests of the Standard Model and searches for New Physics in the quark flavor sector depend on accurate theoretical calculations of decay rates and spectra for rare, flavor-changing processes. The theoretical status and recent developments of techniques allowing such calculations are reviewed. Special attention is paid to the calculation of the $\bar B\to X_s\gamma$ branching ratio, the extraction of the $b$-quark mass from a fit to $\bar B\to X_c\,l\,\bar\nu_l$ moments, and the determination of $|V_{ub}|$ from spectra in the inclusive decay $\bar B\to X_u\,l\,\bar\nu_l$. From a reanalysis of different inclusive distributions the updated average value $|V_{ub}|=(3.98\pm 0.15\pm 0.30)\cdot 10^{-3}$ is derived. Using only the theoretically cleanest channels, we obtain $|V_{ub}|=(3.70\pm 0.15\pm 0.28)\cdot 10^{-3}$.
\end{abstract}

\maketitle

\section{Introduction}

Heavy-quark physics is a corner stone in the ongoing effort to explore the Standard Model of elementary-particle interactions, to determine its parameters with greatest achievable precision, and to search for hints of departures from Standard Model predictions. Thanks to an experimental program at various facilities worldwide that has spanned several decades, accompanied by steady progress in theory, there have been tremendous accomplishments in this field and several important discoveries have been made. While the Standard Model still stands as the most fundamental theoretical structure explaining the fundamental interactions of matter in the Universe, the combined effort of energy-frontier and luminosity-frontier experiments in the coming decade is likely to shed light on the deep questions concerning the physics at the tera-scale -- questions about the origin of mass, the asymmetry between matter and antimatter, and the nature of dark matter.

One of the main challenges in heavy-quark physics is to disentangle the underlying, flavor-changing couplings of the quarks (which in the Standard Model encode all information about CP violation) from the less interesting but all dominating effects of the strong interactions, as described by Quantum Chromodynamics (QCD). Sophisticated theoretical tools going beyond the realm of perturbation theory are required to accomplish percent-level precision in the calculation of decay rates and kinematical distributions for the most interesting processes. Some of the relevant tools and recent developments are discussed in this talk.

\section{Theoretical foundations}

Before discussing specific applications it is important to summarize the theoretical tools we have for calculating decay rates and spectra in heavy-quark physics, and the concepts from which these tools derive. The underlying theme is the separation of short-distance from long-distance physics, which is natural due to the presence of the large mass scale $m_Q\gg\Lambda_{\rm QCD}$, which is far above the scale of nonperturbative hadronic physics. Factorization theorems state that short- and long-distance contributions to a given observable can be separated into Wilson coefficient functions $C_i$ and nonperturbative matrix elements $M_i$. Generically, up to power corrections of the form $(\Lambda_{\rm QCD}/m_Q)^n$ one has
\begin{equation}\label{fact1}
   \mbox{Observable}\sim\sum_i C_i(m_Q,\mu)\,M_i(\mu) + \dots \,.
\end{equation}
Such a factorization formula is useful, since by virtue of it the dependence on the high scale $m_Q$ is calculable, and often the number of matrix elements $M_i$ is smaller than the number of observables that can be expressed in the form shown above. Typically, this reduction is accomplished by means of some symmetry. The nonperturbative matrix elements can then be extracted from data or calculated using theoretical approaches such as lattice QCD or QCD sum rules. The factorization scale $\mu$ in (\ref{fact1}) serves as an auxiliary separator between the domains of short- and long-distance physics. Observables are formally independent of the choice of $\mu$; however, they inherit some residual dependence once the Wilson coefficients $C_i$ are computed at finite order in perturbation theory. The dependence gets weaker as higher orders in the perturbative expansion are included. The $\mu$-independence of the right-hand side of (\ref{fact1}) implies a renormalization-group equation for the functions $C_i$, which can be solved. In the process, large perturbative logarithms of the form $\alpha_s^m\ln^n(m_Q/\mu)$ can be summed to all orders in perturbation theory.

The formal basis of the factorization formula (\ref{fact1}) is the (euclidean) operator product expansion. Important examples of this type of factorization include the effective weak Hamiltonian~\cite{Buchalla:1995vs}
\begin{equation}\label{Heff}
   H_{\rm eff} 
   = \sum_i C_i(m_t,m_W,m_Z,m_H,M_{\rm NP},\mu)\,Q_i(\mu) \,,
\end{equation}
which is the starting point of any calculation in quark flavor physics. In this case the Wilson coefficients $C_i$ contain short-distance physics associated with heavy Standard Model particles such as the top-quark, the electroweak gauge bosons, or the Higgs boson. However, if there exists physics beyond the Standard Model involving some new heavy particles that couple to the Standard Model, then the Wilson coefficients will unavoidably also be sensitive to the masses and couplings of these new particles. An important goal of flavor physics is to search for these kinds of effects. Other important applications of the factorization theorem (\ref{fact1}) include the derivation of heavy-quark symmetry relations for the exclusive semileptonic decays $\bar B\to D^{(*)}l\,\bar\nu_l$ \cite{Isgur:1989ed,Neubert:1993mb}, and the calculation of the total decay rate and differential distributions in the inclusive semileptonic decay $\bar B\to X_c\,l\,\bar\nu_l$ \cite{Bigi:1992su,Bigi:1993fe,Manohar:1993qn}.

Processes involving energetic light partons require a more sophisticated form of the factorization theorem, which generically can be expressed as
\begin{equation}\label{fact2}
   \mbox{Observable}\sim\sum_i H_i(Q,\mu)\,J_i(\sqrt{Q\Lambda},\mu)
   \otimes S_i(\mu) + \dots \,.
\end{equation}
Here $Q\gg\Lambda_{\rm QCD}$ is the hard scale of the process, e.g., a heavy-quark mass or the center-of-mass energy. The hard functions $H_i$ capture virtual effects from quantum fluctuations at the hard scale. The jet functions $J_i$ describe the properties of the emitted collinear particles, whose characteristic virtuality or invariant mass scales as an intermediate scale $\sqrt{Q\Lambda}$ between $Q$ and the QCD scale $\Lambda_{\rm QCD}$. Here we will only be interested in inclusive processes, for which the jet scale is in the perturbative domain ($\sqrt{Q\Lambda}\gg\Lambda_{\rm QCD}$). Finally, the soft functions $S_i$ describe the (nonperturbative)  physics associated with soft radiation in the process. The symbol $\otimes$ implies a convolution, which arises since the jet and soft functions share some common variables, such as some small momentum components of order $\Lambda_{\rm QCD}$. These convolutions require distribution functions (as opposed to hadronic parameters) as hadronic input. Examples of such functions are parton distribution functions, light-cone distribution functions, and fragmentation functions.

The theoretical basis of the factorization theorem (\ref{fact2}) is a generalization of the euclidean operator product expansion to the time-like domain. While in the past factorization theorems such as (\ref{fact2}) were derived using diagrammatic methods \cite{Sterman:1986aj,Catani:1989ne,Korchemsky:1992xv}, an elegant field-theoretic approach is offered by the soft-collinear effective theory developed in \cite{Bauer:2000yr,Bauer:2001yt,Beneke:2002ph,Hill:2002vw}. The interactions of light, energetic particles and the associated emissions of soft and collinear radiation requires a description in terms of nonlocal operators, in which fields live on light-like trajectories following the paths of the energetic partons. As a result, this is a rather complicated effective field theory. However, once its power counting rules have been established, it is possible to classify operators by counting powers of $\Lambda_{\rm QCD}/Q$. 

Many important applications of the factorization theorem (\ref{fact2}) have been discussed in the literature. In jet physics, factorization applies, e.g., to the cross sections for deep-inelastic scattering \cite{Sterman:1986aj,Catani:1989ne,Korchemsky:1992xv,Manohar:2003vb,Becher:2006mr}, Drell-Yan production \cite{Sterman:1986aj,Catani:1989ne,Becher:2007ty}, and to event shapes \cite{Korchemsky:1999kt}. In the realm of heavy-quark physics, formula (\ref{fact2}) applies to the inclusive processes $\bar B\to X_u\,l\,\bar\nu_l$ and $\bar B\to X_s\gamma$, in which case the soft functions $S_i$ are referred to as shape functions \cite{Neubert:1993ch,Neubert:1993um,Bigi:1993ex,Korchemsky:1994jb,Bosch:2004th}. It also applies to exclusive semileptonic processes such as $\bar B\to\pi\,l\,\bar\nu_l$ and $\bar B\to\rho\,l\,\bar\nu_l$ \cite{Bauer:2000yr,Beneke:2000wa}, which have been studied in great detail using light-cone QCD sum rules (for recent work, see e.g.\ \cite{Ball:2004ye,Ball:2004rg}). Also, the QCD factorization approach to nonleptonic processes such as $\bar B\to D^{(*)}\pi$, $\bar B\to\pi K$, and $\bar B\to K^*\gamma$ relies on such a factorization theorem \cite{Beneke:1999br,Beneke:2000ry,Beneke:2001ev,Beneke:2002jn,Beneke:2003zv,Beneke:2001at,Bosch:2001gv}. Some important recent accomplishments in this area have been the calculation of the hard-scattering kernels in the QCD factorization formula for rare hadronic decays of the form $\bar B\to h_1 h_2$ (where $h_i$ are light pseudoscalar or vector mesons) at $O(\alpha_s^2)$ \cite{Beneke:2005vv,Beneke:2006mk,Bell:2007tv}, the proof of factorization for the decay $\bar B\to K^*\gamma$ \cite{Becher:2005fg}, and the calculation of part of the $O(\alpha_s^2)$ corrections for this process~\cite{Ali:2007sj}.

Any systematic theoretical approach has limitations, an understanding of which is necessary in order to gauge the accuracy of calculations and, more generally, understand what can and what cannot be calculated in a model-independent way. An obvious practical limitation of any calculation is the need to truncate the perturbative expansion of the perturbative coefficients $C_i$ in (\ref{fact1}), and $H_i$, $J_i$ in (\ref{fact2}). State of the art is to compute these objects at next-to-next-to-leading order (NNLO) in perturbation theory. Even if it was possible to calculate the perturbative functions to all orders (and of course we do not have the tools to accomplish this!), then the asymptotic nature of perturbative expansions in quantum field theory would imply an irreducible ambiguity in the precision with which these functions can be determined, an ambiguity that scales like a power of $\Lambda_{\rm QCD}/Q$. In order to do better, it is necessary to include power-suppressed corrections in the calculation. In some cases such as the inclusive decays $\bar B\to X_u\,l\,\bar\nu$ and $\bar B\to X_s\gamma$, a factorization theorem such as (\ref{fact2}) can be shown to hold at any order in the expansion in $\Lambda_{\rm QCD}/m_b$ \cite{Lee:2004ja,Bosch:2004cb,Beneke:2004in}; however, starting already at first order in $\Lambda_{\rm QCD}/m_b$ the number of unknown soft functions exceeds the number of observables. In other cases, such as the nonleptonic decays $\bar B\to\pi K$ and $\bar B\to\pi\pi$, factorization breaks down beyond the leading power in $\Lambda_{\rm QCD}/m_b$ \cite{Beneke:1999br,Beneke:2000ry,Beneke:2001ev}. It follows from these remarks that observables very sensitive to higher-order perturbative effects or power corrections cannot be predicted with high accuracy. Well-known examples include some direct CP asymmetries for the decays $\bar B\to\pi K$ and $\bar B\to\pi\pi$, and the branching fractions for some ``color-suppressed" decays such as $\bar B\to\pi^0\pi^0$. The fact that calculations are rather uncertain in these cases does not indicate a failure of the factorization framework (which is rooted in quantum field theory and as such cannot be questioned), but rather that in some cases the expansion in powers of $\alpha_s$ and $\Lambda_{\rm QCD}/m_b$ does not converge well.

Before turning to phenomenological applications we should mention an ongoing theoretical debate on the question of the breakdown of factorization beyond the leading power in heavy-to-light transitions such as the $\bar B\to\pi$ form factor or hadronic processes such as $\bar B\to\pi\pi$. It is well known that a perturbative analysis of the form factor is plagued by ``endpoint divergent" convolution integrals \cite{Szczepaniak:1990dt,Burdman:1992hg,Bauer:2002aj,Beneke:2003pa,Lange:2003pk}. Power corrections to the factorization formula for $\bar B\to\pi\pi$, which arise from processes such as ``weak annihilation", give rise to similar endpoint divergences \cite{Beneke:2000ry,Beneke:2001ev}. It has recently been claimed that these divergences can be removed, and factorization be established, in a systematic way using so-called ``0-bin" subtractions combined with rapidity factorization \cite{Manohar:2006nz}. Unfortunately this claim has not been supported by detailed calculations, and in my view it is unlikely that the problem can be solved following the approach described in that paper. Nevertheless, based on these results it has been argued that weak annihilation amplitudes \cite{Arnesen:2006vb} and power-suppressed penguin amplitudes \cite{Jain:2007dy} should be real to a good approximation, i.e., free of large soft rescattering phases. If true, this would have important implications for phenomenology. Finally, the prediction of the QCD factorization approach that charm-penguin contributions to nonleptonic decay amplitudes factorize \cite{Beneke:1999br,Beneke:2000ry,Beneke:2001ev} has been challenged in \cite{Bauer:2004tj}, and a rebuttal has been given in \cite{Beneke:2004bn}. More theoretical work on this important issue would be desirable.

The remainder of this talk will focus on three important applications of heavy-quark theory: the calculation of the partial inclusive $\bar B\to X_s\gamma$ branching ratio at NNLO, the extraction of the $b$-quark mass from moments of inclusive decay spectra, and the determination of $|V_{ub}|$ from inclusive $\bar B\to X_u\,l\,\bar\nu_l$ decay distributions. New Physics implications of $\bar B\to X_s\gamma$ decay have been discussed by G.~Isidori and S.~Heinemeyer at this conference, while determinations of $|V_{ub}|$ from exclusive decays have been covered in the talks of E.~Barberio and C.~Davis. For a lack of time, this is a very limited selection of topics from a vast range of possibilities. My apologies go to all those authors whose important contributions are omitted here.

\section{The $\bar B\to X_s\gamma$ branching ratio}

The decay $\bar B\to X_s\gamma$ is the prototype of a flavor-changing neutral current process. In the Standard Model it is mediated by loops involving heavy top quarks and $W$ bosons, and it is expected that yet unknown heavy particles, if they exist at the TeV scale, could give contributions to the decay rate not much below the Standard Model level. A precise study of this process, both experimentally and theoretically, can thus serve as a precision test of the flavor sector and place constraints on the parameters of New Physics models. Indeed, it is by now broadly appreciated that the constraints on model building from processes such as $\bar B\to X_s\gamma$ can be as valuable as constraints from electroweak precision observables, astrophysical observations, and other low-energy observables such as electric dipole moments and the anomalous magnetic moment of the muon.

About one year ago, the calculation of most NNLO contributions (all that are believed to be significant) to the $\bar B\to X_s\gamma$ decay rate has been completed, and a first prediction for the branching ratio has been derived in a paper by Misiak at 16 coworkers \cite{Misiak:2006zs}. This has been a heroic effort, which would not have been possible without a large collaborative approach. At the same time, a dedicated, model-independent analysis of the effect of a cut on the photon energy, $E_\gamma>E_0$ with $E_0$ in the range between 1.8 and 2\,GeV as applied in the experimental analyses, has been performed \cite{Becher:2005pd,Becher:2006qw,Becher:2006pu}. It is conventional (yet somewhat artificial) to write the branching ratio as $\mbox{Br}(E_\gamma>E_0)=\mbox{Br(tot)}\,F(E_0)$, where the function $F(E_0)$ describes the effect of the cut, while Br(tot) is the ``total" branching ratio (which in practice refers to a very low cut $E_\gamma>1$\,GeV, since the total branching ratio does not exist). 

We begin with a brief review of the NNLO calculation of the total branching ratio. The first task was to compute the Wilson coefficients in the effective weak Hamiltonian (\ref{Heff}) at NNLO. This required two-loop matching calculations for the coefficients $C_{1-6}$ at the weak scale \cite{Bobeth:1999mk} and three-loop matching calculations for the coefficients $C_{7\gamma}$ and $C_{8g}$ \cite{Misiak:2004ew}. These matching conditions are the place where New Physics contributions would enter the theoretical analysis \cite{Bobeth:1999ww}. Next, the Wilson coefficients have to be evolved down to a scale $\mu\sim m_b$ using the renormalization group. As input, one needs the three-loop anomalous dimension matrices for the operators $Q_{1-6}$ \cite{Gorbahn:2004my} and $Q_{7\gamma}$, $Q_{8g}$ \cite{Gorbahn:2005sa}, as well as the four-loop anomalous dimensions accounting for the mixing of $Q_{1-6}$ into $Q_{7\gamma}$, $Q_{8g}$ \cite{Czakon:2006ss}. With these results at hand, the second task was to compute the matrix elements of the various operators at NNLO. The full two-loop corrections to the matrix element of the leading dipole operator $Q_{7\gamma}$ were obtained in \cite{Melnikov:2005bx,Blokland:2005uk,Asatrian:2006ph,Asatrian:2006sm}. For the required three-loop matrix elements of the operators $Q_{1,2}$ only an extrapolation in the mass of the charm-quark is available based on calculations performed in the artificial limit $m_c\gg m_b/2$ \cite{Misiak:2006ab}. One of the dominant uncertainties in the prediction results from this extrapolation. The remaining matrix elements are known at $O(\beta_0\alpha_s^2)$ \cite{Ligeti:1999ea,Bieri:2003ue}. The resulting prediction for the total branching ratio in the Standard Model is \cite{Misiak:2006zs}
\begin{equation}\label{bsgrate}
   \mbox{Br}(E_\gamma>1\,\mbox{GeV}) = (3.27\pm 0.23)\cdot 10^{-4} \,.
\end{equation}
The combined theoretical uncertainty results from a 3\% error due to higher-order perturbative corrections, a 3\% error in the extrapolation in the charm-quark mass, another 3\% uncertainty from parameter variations (from $m_t$, $m_c$, $\alpha_s$ etc.), and finally a 5\% error assigned to account for nonperturbative power corrections to the total rate. 

The result (\ref{bsgrate}) is a mile stone in the history of heavy-flavor theory. Note that the largest contribution to the theoretical uncertainty is now due to our lack of control over nonperturbative power corrections. Unfortunately, it will be very difficult if not impossible to reduce this error with current technology. A recent analysis has identified a new class of nonlocal power corrections to the total decay rate, which are described in terms of matrix elements of trilocal light-cone operators \cite{Lee:2006wn}. Naive model estimates of these effects point to a small reduction of the total rate with an uncertainty of 5\%. However, this number is only a guess, and it is difficult to substantiate it. There are other, similar types of nonlocal contributions, which are currently being analyzed \cite{inprep}. 

The presence of a cut $E_\gamma>E_0$ on the photon energy, which is applied in order to reduce the background in all measurements of the $\bar B\to X_s\gamma$ branching ratio to date, leads to significant complications in the theoretical analysis. The reason is that it introduces a sensitivity to the low scale $\Delta=m_b-2E_0\approx 1$\,GeV, which is barely in the perturbative domain. One thus needs to deal with a complicated many-scale problem characterized by the hierarchy $m_b>\sqrt{m_b\Delta}>\Delta>\Lambda_{\rm QCD}$. A systematic framework for this situation is the multi-scale operator product expansion developed in \cite{Neubert:2004dd,Neubert:2005nt}. In this case the theoretical accuracy that can be achieved is {\em not\/} set by size of the expansion parameters $\alpha_s(m_b)\approx 0.2$ and $\Lambda_{\rm QCD}/m_b\approx 0.1$, but rather by $\alpha_s(\Delta)\approx 0.4$ and $\Lambda_{\rm QCD}/\Delta\approx 0.5$ (though first-order corrections in this ratio are absent). A detailed analysis at NNLO comes to the conclusion that \cite{Becher:2006pu}
\begin{equation}\label{FE0}
   F(1.6\,\mbox{GeV}) = 0.93_{\,-0.06}^{\,+0.04} \,,
\end{equation}
which tends to be a stronger suppression than that predicted by naive model estimates \cite{Buchmuller:2005zv}. The above number includes a perturbative uncertainty of ${}_{\,-5}^{\,+3}\,\%$, parameter dependences of 2\%, and an error from power corrections of also 2\%.

Combining the results (\ref{bsgrate}) and (\ref{FE0}), one obtains the NNLO prediction \cite{Becher:2006pu}
\begin{equation}\label{bsgfinal}
   \mbox{Br}(E_\gamma>1.6\,\mbox{GeV}) 
   = (2.98\pm 0.26)\cdot 10^{-4} \,.
\end{equation}
It is consistent within errors with the result $(3.15\pm 0.23)\cdot 10^{-4}$ quoted in \cite{Misiak:2006zs}, which is based on a naive perturbative expansion in $\alpha_s(m_b)$. For completeness we also quote the number $(3.49\pm 0.49)\cdot 10^{-4}$ obtained in \cite{Andersen:2006hr} using a renormalon-inspired model for the shape function in $\bar B\to X_s\gamma$ decay. Note that the difference with respect to (\ref{bsgfinal}) is primarily due to a larger total branching ratio obtained in this scheme, not so much due to the treatment of cut-related effects.

The theoretical result (\ref{bsgfinal}) for the branching ratio is about 1.4 standard deviations lower than the current world-average experimental value $\mbox{Br}(E_\gamma>1.6\,\mbox{GeV})=(3.55\pm 0.26)\cdot 10^{-4}$ \cite{Barberio:2007cr}, indicating that there is some room for possible New Physics effects. We emphasize, however, that to date no actual measurement has been performed with a cut as low as 1.6\,GeV. Indeed, the quoted experimental number involves a model-dependent extrapolation to low energy. In the future, it would be desirable if experiment and theory would be confronted at a value of $E_0$ that can indeed be achieved experimentally. The lowest such value at present is $E_0=1.8$\,GeV.

\section{Extraction of $m_b$ from a fit to $\bar B\to X_c\,l\,\bar\nu_l$ moments}

The operator product expansion allows for a model-independent calculation of moments of decay distributions in the inclusive semileptonic process $\bar B\to X_c\,l\,\bar\nu_l$ \cite{Bigi:1993fe,Manohar:1993qn}. Hadronic physics is encoded in a few parameters defined in terms of forward $B$-meson matrix elements of local operators. They are called $\mu_\pi^2$ (or $\lambda_1$), $\mu_G^2$ (or $\lambda_2$), etc. The most important parameters in the prediction are, however, the heavy-quark masses $m_b$ and $m_c$. The only assumption underlying the theoretical calculations is that of quark-hadron duality. It is believed to be reliable for the relevant energy release $\Delta E\sim m_B-m_D$.

Traditionally, the parameters $\{|V_{cb}|, m_b, m_c, \mu_\pi^2, \mu_G^2\}$ are extracted from a global fit to experimental data on moments of the lepton energy and invariant hadronic mass spectra in $\bar B\to X_c\,l\,\bar\nu_l$ decay and of the photon energy spectrum in $\bar B\to X_s\gamma$ decay. The experimental data include measurements reported by the BaBar, Belle, CLEO, CDF, and DELPHI collaborations \cite{Barberio:2007cr}, while the theoretical calculations are based on formulae derived in \cite{Bauer:2002sh,Bigi:2003zg,Gambino:2004qm,Bauer:2004ve}. The global fit takes account of the strong correlations between the various quantities. The status of the theoretical calculations is such that the leading terms in the operator product expansion for the moments are know at $O(\beta_0\alpha_s^2)$ but not yet at $O(\alpha_s^2)$. Power-suppressed contributions are only known at tree level. It will be important in the future to increase the accuracy of the calculations. The technology for obtaining the complete $O(\alpha_s^2)$ corrections to the leading term exists \cite{Anastasiou:2005pn}, and it is expected that results will soon be published. Also, the $O(\alpha_s)$ corrections to the leading power-suppressed effects, those proportional to the parameters $\mu_\pi^2$ and $\mu_G^2$, are currently being calculated. Results have already been published for the terms proportional to $\mu_\pi^2$ \cite{Becher:2007tk}, but they have not yet been included in the fits. 
 
\begin{table*}
\caption{\label{tab:1}
Translation of two different $m_b$ values from the kinetic scheme to other schemes, at different orders in perturbation theory. The errors in columns two to four indicate the scale dependence ($m_b/2<\mu<2m_b$). The last column shows some reference values for comparison.}
\vspace{0.1cm}
\begin{center}
\begin{tabular}{lcccccl}
\hline
$m_b^{\rm kin}=4613$\,MeV && $O(\alpha_s)$ & $O(\beta_0\alpha_s^2)$
 & $O(\alpha_s^2)$ && Reference values \\
\hline\\[-0.4cm]
$\overline{m}_b(\overline{m}_b)$ [MeV]
 && $4317_{\,-88}^{\,+54}$
 & $4159_{\,-96}^{\,+67}$
 & $4195_{\,-72}^{\,+55}$
 && $4164\pm 25$ ($e^+ e^-\to\mbox{hadrons}$ \cite{Kuhn:2007vp}) \\
$m_b^{\rm 1S}$ [MeV]
 && $4698_{\,-\phantom{8}8}^{\,+\phantom{8}6}$
 & $4745_{\,-\phantom{8}9}^{\,+\phantom{8}9}$
 & $4742_{\,-\phantom{8}9}^{\,+13}$ 
 && $4701\pm 30$ (fit in 1S scheme \cite{Barberio:2007cr}) \\
$m_b^{\rm SF}$ [MeV]
 && $4609_{\,-\phantom{8}2}^{\,+\phantom{8}1}$
 & $4584_{\,-16}^{\,+\phantom{8}8}$
 & $4643_{\,-11}^{\,+24}$ 
 && $4630\pm 60$ (used by HFAG \cite{Barberio:2007cr}) \\[0.1cm]
 \hline
$m_b^{\rm kin}=4677$\,MeV &&&&&& \\
\hline\\[-0.4cm]
$\overline{m}_b(\overline{m}_b)$ [MeV]
 && $4377_{\,-89}^{\,+55}$
 & $4217_{\,-97}^{\,+67}$
 & $4253_{\,-72}^{\,+55}$
 && $4164\pm 25$ ($e^+ e^-\to\mbox{hadrons}$ \cite{Kuhn:2007vp}) \\
$m_b^{\rm 1S}$ [MeV]
 && $4761_{\,-\phantom{8}8}^{\,+\phantom{8}6}$
 & $4807_{\,-\phantom{8}9}^{\,+\phantom{8}8}$
 & $4804_{\,-\phantom{8}9}^{\,+12}$ 
 && $4751\pm 58$ (fit in 1S scheme \cite{Barberio:2007cr}) \\
$m_b^{\rm SF}$ [MeV]
 && $4671_{\,-\phantom{8}2}^{\,+\phantom{8}1}$
 & $4648_{\,-16}^{\,+\phantom{8}8}$
 & $4706_{\,-11}^{\,+24}$ 
 && $4707\pm 56$ (HFAG update for LP07) \\[0.1cm]
\hline
\end{tabular}
\end{center}
\end{table*}

The fit strategy is as follows: A set of experimental results, including errors and correlations, is fitted to a set of theoretical equations derived using the operator product expansion. The fit is performed in the space of parameters $\{|V_{cb}|, m_b, m_c, \mu_\pi^2, \mu_G^2\}$ mentioned above. Besides $|V_{cb}|$, these parameters must be defined in a particular renormalization and subtraction scheme. The most popular schemes are the kinetic scheme \cite{Benson:2003kp}, the 1S scheme \cite{Hoang:1998hm}, and the shape-function scheme \cite{Bosch:2004th,Neubert:2004sp}.  While in principle it is merely a matter of choice which of these schemes one prefers, fitting in scheme~A and converting the results to scheme~B might give slightly different results from fitting directly in scheme~B, if truncated perturbative expressions are employed. This fact should be taken into account when deriving estimates for the theoretical uncertainties; specifically, the quoted uncertainties should not be less than the uncertainties inherent in the scheme translations. To illustrate this point, we quote results derived by the Heavy Flavor Averaging Group (HFAG) for this conference. They obtain $m_b^{\rm kin}=(4.613\pm 0.035)$\,GeV from a fit performed in the kinetic scheme and $m_b^{\rm 1S}=(4.701\pm 0.030)$\,GeV from a fit performed in the 1S scheme \cite{Barberio:2007cr}. On the other hand, translating the first number from the kinetic to the 1S scheme at $O(\beta_0\alpha_s^2)$ would give $m_b^{\rm 1S~from~kin}=(4.745\pm 0.036)$\,GeV, 
which is about 1.5 standard deviations higher than the value obtained from the fit in the 1S scheme. This indicates that the uncertainty is larger than the 30\,MeV quoted from the fit.

Table~\ref{tab:1} collects results obtained by translating the central values $m_b^{\rm kin}=4.613$\,GeV (upper portion) and $m_b^{\rm kin}=4.677$\,GeV (lower portion) from the kinetic scheme into other schemes ($\overline{\rm MS}$, 1S, and shape-function scheme). The formulae translating from one scheme to another involve perturbative expansions, and following common practice we choose to evaluate the coupling $\alpha_s$ in these expansions at a scale $\mu$ varied between $m_b/2$ and $2m_b$. This results in the quoted uncertainties. A few comments are worth mentioning: First, note that the scale uncertainty from the conversion into the $\overline{\rm MS}$ scheme remains significant even at two-loop order. This unfortunate fact prevents us from using the very precise value for $\overline{m}_b(\overline{m}_b)$ recently obtained from an analysis of the total $e^+ e^-\to\mbox{hadrons}$ cross section in the threshold region \cite{Kuhn:2007vp} to derive similarly precise values for the $b$-quark mass in other schemes. Next, note that the value of $m_b$ at NLO in the 1S scheme has a tiny scale uncertainty, yet the value obtained at NNLO differs from the NLO value by more than 40\,MeV. This illustrates that scale variation can sometimes underestimate the true theoretical uncertainty. Finally, note that for the translation into the shape-function scheme two-loop corrections not associated with running coupling effects appear to be important. It would therefore be desirable to include the full $O(\alpha_s^2)$ corrections in the moment fit.

\begin{figure*}
\begin{center}
\epsfig{file=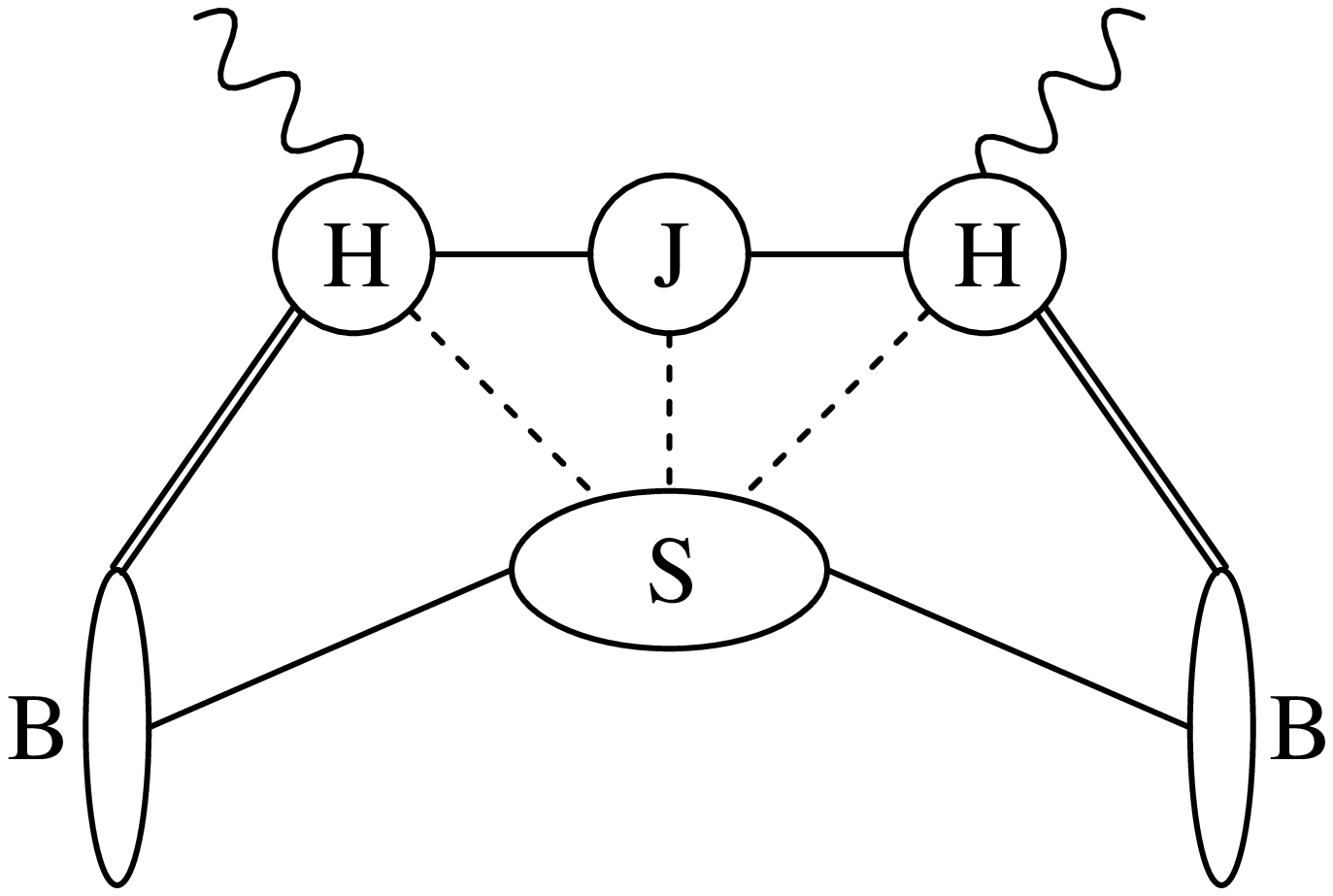,width=4.5cm}\qquad
\epsfig{file=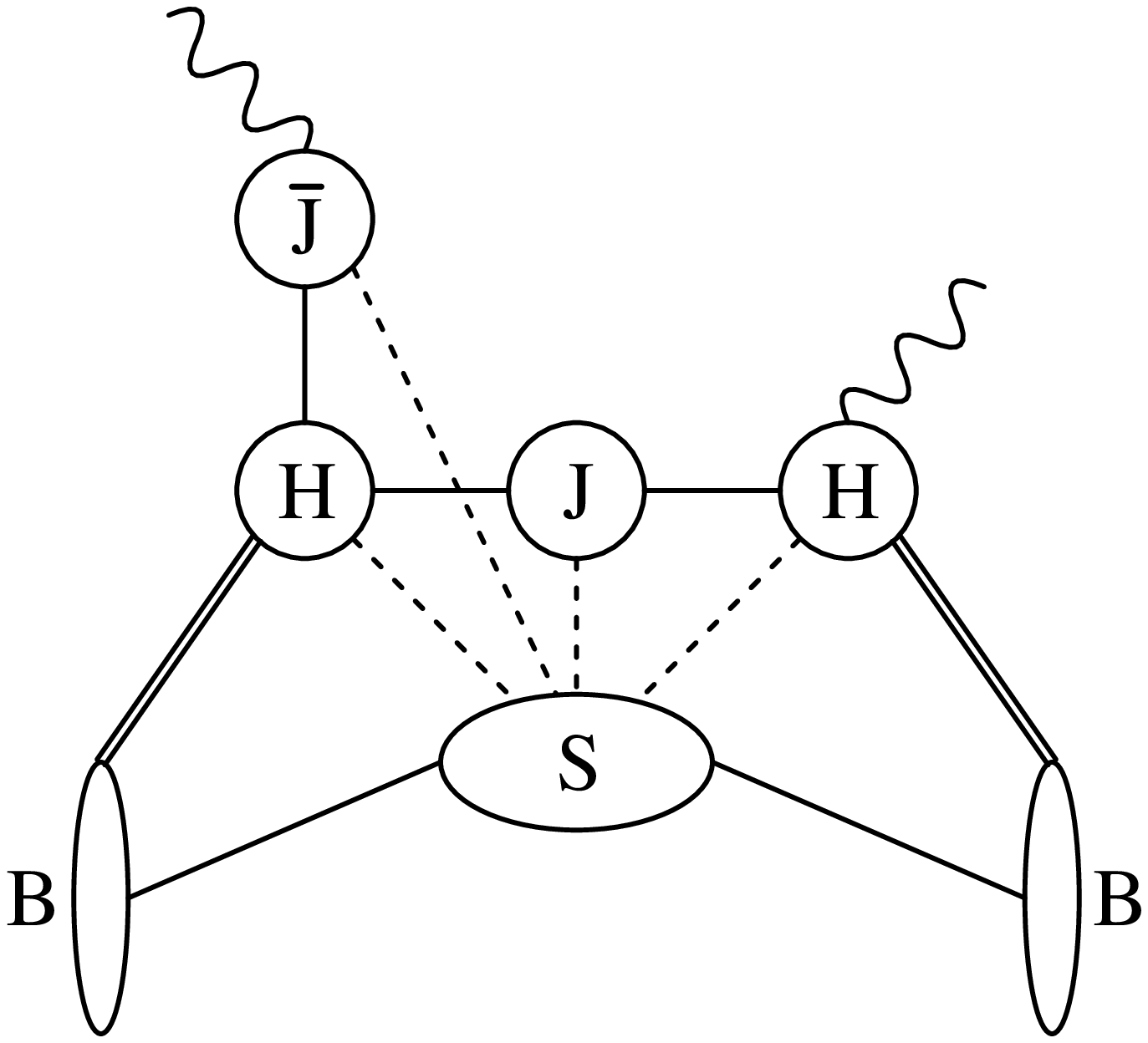,width=4.5cm}\qquad
\epsfig{file=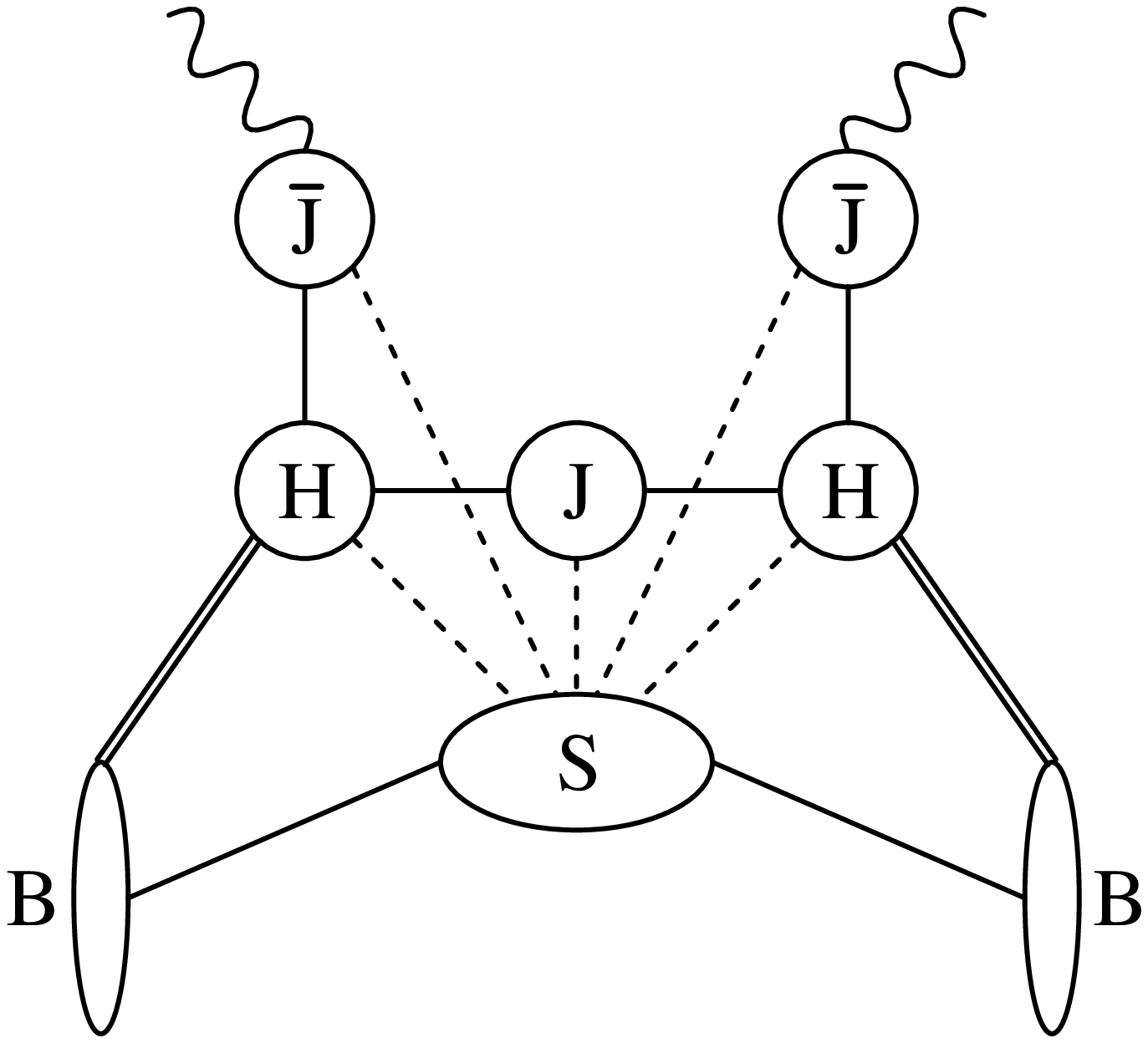,width=4.5cm}
\caption{\label{fig:theorem}
Graphical illustration of the QCD factorization theorem for $\bar B\to X_s\gamma$ decay in the endpoint region \cite{inprep}. The dashed lines represent soft interactions, which must be factored off the remaining building blocks to prove factorization.}
\end{center}
\end{figure*}

A serious problem with the way the fit is currently performed by HFAG is that besides the moments of semileptonic spectra in $\bar B\to X_c\,l\,\bar\nu_l$ decay also moments of the $\bar B\to X_s\gamma$ photon energy spectrum are used. Based on what we know about $\bar B\to X_s\gamma$ decay, this treatment can no longer be justified theoretically! Whereas a clean theoretical approach to the $\bar B\to X_c\,l\,\bar\nu_l$ moments is provided by the operator product expansion, the theoretical basis for the calculation of the $\bar B\to X_s\gamma$ moments is far more complicated and subject to uncontrollable theoretical uncertainties. First, it is unquestionable that the existing measurements of the $\bar B\to X_s\gamma$ moments are performed in a kinematic region where shape-function effects are still important \cite{Kagan:1998ym}. To correct for these effects, a (shape-function) model-dependent ``bias correction" is applied to the data \cite{Benson:2004sg}. More generally, it has now been established that as soon as operators other than the dipole operator $Q_{7\gamma}$ are not ignored, then {\em there is no local operator product expansion\/} for the $\bar B\to X_s\gamma$ decay rate and moments, even outside the shape-function region \cite{Lee:2006wn,inprep}. Indeed, a novel factorization formula can be derived for the photon energy spectrum, which is graphically represented in Figure~\ref{fig:theorem}. Besides a ``direct photon" contribution (first term), there appear ``single resolved" and ``double resolved photon" contributions, which are impossible to calculate with present tools. Their impact on the moments of the photon spectrum is currently under investigation. 

\begin{figure}
\begin{center}
\epsfig{file=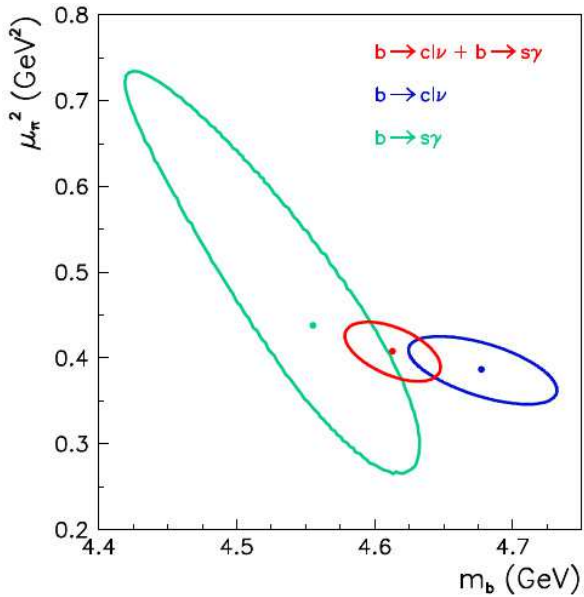,width=0.9\columnwidth}
\caption{\label{fig:BFplot}
Results for the parameters $m_b$ and $\mu_\pi^2$ (defined in the kinetic scheme) obtained from moment fits to $\bar B\to X_c\,l\,\bar\nu_l$ (blue), $\bar B\to X_s\gamma$ (green), and both (red). The plot is an update of a figure in \cite{Buchmuller:2005zv} (courtesy of HFAG).}
\end{center}
\end{figure}

When the goal is to extract heavy-quark parameters such as $m_b$ in a model-independent way, one should thus not include the $\bar B\to X_s\gamma$ moments in the fit. An analysis based only on the $\bar B\to X_c\,l\,\bar\nu_l$ moments leads to $m_b^{\rm kin}=(4.677\pm 0.053)$\,GeV in the kinetic scheme and $m_b^{\rm 1S}=(4.751\pm 0.058)$\,GeV in the 1S scheme \cite{Barberio:2007cr}. The increase of the uncertainties can be understood by observing that in the combined fit there was s slight tension between the results from the $\bar B\to X_c\,l\,\bar\nu_l$ and $\bar B\to X_s\gamma$ moments, as illustrated in Figure~\ref{fig:BFplot}. The increased errors shown above more properly reflect the true uncertainties. These errors will be reduced (perhaps to 30--40\,MeV) once the $O(\alpha_s^2)$ and $O(\alpha_s/m_b^2)$ corrections will be included in the fit.

When the new values for $m_b$ are translated into the shape-function scheme, one obtains $m_b^{\rm SF}=(4.707\pm 0.056)$\,GeV, which is larger than the value $(4.63\pm 0.06)$\,GeV used so far by HFAG. This observation has important implications for the determination of $|V_{ub}|$.

\section{Inclusive determination of $|V_{ub}|$}

The theoretical basis for the analysis of differential decay distributions in the inclusive semileptonic process $\bar B\to X_u\,l\,\bar\nu_l$ is a factorization formula of the form (\ref{fact2}), in which the soft functions $S_i$ are generalized parton distribution functions for the $B$ meson called shape functions \cite{Neubert:1993ch,Neubert:1993um,Bigi:1993ex,Korchemsky:1994jb,Bosch:2004th}. The leading-order shape function is a universal, process-independent quantity, which also (to a large extent) determines the shape of the $\bar B\to X_s\gamma$ photon energy spectrum. The strategy is therefore to extract the shape function from the $\bar B\to X_s\gamma$ photon spectrum and then use this information to predict $\bar B\to X_u\,l\,\bar\nu_l$ decay distributions \cite{Bosch:2004th,Bosch:2004bt,Lange:2005yw}. The functional form of the shape function is constrained by moment relations, which relate weighted integrals over the shape function to the heavy-quark parameters $m_b$, $\mu_\pi^2$ etc.\ extracted from the $\bar B\to X_c\,l\,\bar\nu_l$ moment fit. An alternative is to employ shape-function independent relations between weighted $\bar B\to X_s\gamma$ and $\bar B\to X_c\,l\,\bar\nu_l$ spectra \cite{Neubert:1993um,Leibovich:1999xf,Leibovich:2000ey,Lange:2005qn,Lange:2005xz}. Both approaches are equivalent; yet, not all predictions that have been obtained using the shape-function independent relations are up to the standard of present-day calculations.

The elimination of background events from $\bar B\to X_c\,l\,\bar\nu_l$, in which the charm quark is misidentified, is accomplished by means of different experimental cuts. The most common ones are a cut $M_X<m_D$ on the hadronic invariant mass of the final state, a cut $E_l>(m_B^2-m_D^2)/2m_B$ on the energy of the charged lepton, a cut $q^2>(m_B-m_D)^2$ on the invariant mass squared of the lepton-neutrino pair, or a cut $P_+<m_D^2/m_B$ on the plus component of the total momentum of the final-state hadrons. All are designed such that they eliminate charm background while keeping some portion of the signal events. From a theoretical perspective the ideal cut is that on hadronic invariant mass, followed by the $P_+$ cut. They keep more than 50\% of the signal events. Cutting on lepton energy or leptonic mass is far less efficient.

\begin{table*}
\caption{\label{tab:Vub}
Compilation of $|V_{ub}|$ values obtained using different partial $\bar B\to X_u\,l\,\bar\nu_l$ decay rates, analyzed using the BLNP approach \cite{Bosch:2004th,Lange:2005yw}. The first column of $|V_{ub}|$ values is obtained by extracting the $b$-quark mass from the combined fit to $\bar B\to X_c\,l\,\bar\nu_l$ and $\bar B\to X_s\gamma$ moments \cite{Barberio:2007cr}. The second column is obtained by only using the theoretically clean information from $\bar B\to X_c\,l\,\bar\nu_l$ moments (courtesy of HFAG).}
\vspace{0.1cm}
\begin{center}
\begin{tabular}{lrcccc}
\hline
Method & $a$~~ && $|V_{ub}|~[10^{-3}]$ (old)
 && $|V_{ub}|~[10^{-3}]$ (new) \\
\hspace{4.0cm} $m_b^{\rm SF}$ [GeV]: &&& $4.63\pm 0.06$ && $4.71\pm 0.06$ \\
\hline\\[-0.4cm]
CLEO, $E_l>2.1$\,GeV & 14.4
 && $3.91\pm 0.46\pm 0.44$ && $3.52\pm 0.41\pm 0.35$ \\
Belle, $E_l>1.9$\,GeV & 9.5
 && $4.67\pm 0.43\pm 0.36$ && $4.35\pm 0.40\pm 0.33$ \\
BaBar, $E_l>2.0$\,GeV & 11.4
 && $4.23\pm 0.24\pm 0.39$ && $3.89\pm 0.22\pm 0.33$ \\
BaBar, $E_l>2.0$\,GeV, $s_h^{\rm max}<3.5$\,GeV$^2$ & 13.8
 && $4.37\pm 0.29\pm 0.49$ && $3.94\pm 0.27\pm 0.39$ \\
Belle, $M_X<1.7$\,GeV & 10.5
 && $3.92\pm 0.26\pm 0.32$ && $3.66\pm 0.24\pm 0.27$ \\
BaBar, $M_X<1.55$\,GeV & 13.6
 && $4.09\pm 0.20\pm 0.39$ && $3.74\pm 0.18\pm 0.31$ \\
Belle, $M_X<1.7$\,GeV, $q^2>8$\,GeV$^2$ & 9.3
 && $4.23\pm 0.45\pm 0.36$ && $3.97\pm 0.42\pm 0.31$ \\[0.1cm]
Average &&& $4.31\pm 0.17\pm 0.35$ && $3.98\pm 0.15\pm 0.30$ \\[0.1cm]
\hline
\end{tabular}
\end{center}
\end{table*}

Until this summer, the most complete theoretical analysis of inclusive $\bar B\to X_u\,l\,\bar\nu_l$ spectra was based on calculations described in \cite{Bosch:2004th,Lange:2005yw} and referred to as the BLNP approach. It includes complete perturbative calculations at NLO with Sudakov resummation, subleading shape functions at tree level, and kinematical power corrections at $O(\alpha_s)$. The calculation of $O(\alpha_s^2)$ corrections is in progress and could have a significant impact \cite{inprep2}. An alternative scheme called ``Dressed Gluon Exponentiation" (DGE) \cite{Gardi:2004ia,Andersen:2005bj,Andersen:2005mj} employs a renormalon-inspired model for the leading shape function, which is less flexible in its functional form that the forms used by BLNP. Also, no attempt is made to include subleading shape functions, which among other things means that the predictions for more inclusive partial rates would not be in accord with the operator product expansion beyond the leading power in $\Lambda_{\rm QCD}/m_b$. The DGE model is therefore less rigorous than the BLNP approach, even though it leads to numerical results that are compatible with those of BLNP. Very recently, Gambino et al.\ (GGOU) \cite{Gambino:2007rp} have produced a code for $\bar B\to X_u\,l\,\bar\nu_l$ spectra that is comparable in its completeness with that of BLNP. Working in the kinetic scheme, these authors use moment constraints to model the subleading shape functions differently for the various structures in the hadronic tensor. They include the recently calculated $O(\beta_0\alpha_s^2)$ corrections to the decay rates \cite{Gambino:2006wk}, which were not yet available for BLNP. On the other hand, no attempt is made to resum Sudakov logarithms. The numerical results obtained using the GGOU code are consistent with those of BLNP in both their central values and error estimates.

\begin{figure}
\begin{center}
\epsfig{file=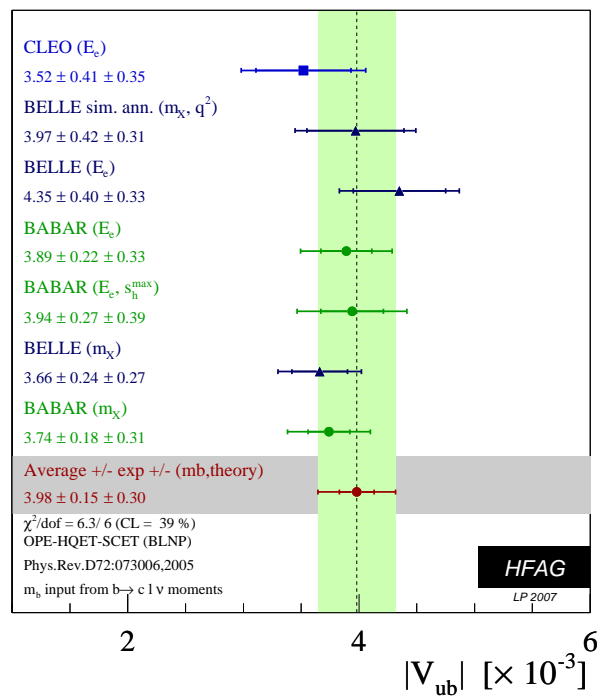,width=0.9\columnwidth}
\caption{\label{fig:Vub}
Compilation of $|V_{ub}|$ determinations from inclusive spectra in $\bar B\to X_u\,l\,\bar\nu_l$ decay, using $m_b^{\rm SF}=(4.71\pm 0.06)$\,GeV for the $b$-quark mass, as determined using only model-independent information for $\bar B\to X_c\,l\,\bar\nu_l$ moments (courtesy of HFAG).}
\end{center}
\end{figure}

In Table~\ref{tab:Vub} we compile results for $|V_{ub}|$ obtained from various various experimental cuts. The data have been analyzed using the BLNP approach. It is crucial to realize that these results depend very sensitively on the shape function used in the analysis, and in particular on its first moment, which is related to the $b$-quark mass. The coefficient $a$ given for each cut indicates how the partial rate scales with $m_b$. The extracted value of $|V_{ub}|$ thus scales like $(m_b)^{-a/2}$. The third column in the table shows results reported previously by HFAG, which are based on using the old value $m_b^{\rm SF}=(4.63\pm 0.06)$\,GeV for the $b$-quark mass. Following my request, HFAG has now redone the analysis using the new value $m_b^{\rm SF}=(4.71\pm 0.06)$\,GeV obtained without including the $\bar B\to X_s\gamma$ moments in the global fit. The new combined average value is
\begin{equation}
   |V_{ub}| = (3.98\pm 0.15\pm 0.30)\cdot 10^{-3} \,.
\end{equation}
It is significantly lower than the previous value. A graphical representation of the individual determinations is shown in Figure~\ref{fig:Vub}. Note that there is rather good agreement among the different measurements. From a theoretical perspective, one would give preference to the extractions based on the most efficient cuts, such as the cuts on $M_X$ or $P_+$ (not reported in the table). For these quark-hadron duality is expected to work best and, more importantly, the potential contamination from weak annihilation effects is minimal. Averaging the BaBar and Belle results derived form the $M_X$ cut, my personal ``best guess" for $|V_{ub}|$ is
\begin{equation}
   |V_{ub}| = (3.70\pm 0.15\pm 0.28)\cdot 10^{-3} \,.
\end{equation}

The most important conclusion from this discussion is this: When only model-independent information is used to determine the $b$-quark mass and thereby derive constraints on the leading shape function, then the value extracted for $|V_{ub}|$ from inclusive decays is in rather good agreement with that derived from exclusive decays. For instance, recent analyses of the $q^2$ distribution in $\bar B\to\pi\,l\,\bar\nu_l$ decays using unquenched lattice QCD \cite{Okamoto:2004xg,Dalgic:2006dt} or light-cone QCD sum rules \cite{Ball:2004ye} yield values in the range between $(3.3\mbox{--}3.6)\cdot 10^{-3}$, with uncertainties of about ${}_{-0.4}^{+0.6}\cdot 10^{-3}$ \cite{Barberio:2007cr}. In my view there is no obstacle to averaging the ``inclusive" and ``exclusive" results to obtain a single value for $|V_{ub}|$, but this is best done by HFAG.

\section{Conclusions}

An important theoretical challenge in heavy-flavor physics is to disentangle strong-interaction effects from the underlying weak-interaction transitions, which probe the flavor structure of the Standard Model and measure fundamental parameters such as the elements of the CKM matrix. Tremendous progress in theory has been made based on heavy-quark expansions and factorization techniques combined with effective field-theory methods. Thanks to these developments, and to the hard work of many theorists, we now have accurate predictions, including high-order perturbative effects and nonperturbative power corrections, for a number of important decay processes, including in particular the inclusive processes $\bar B\to X_s\gamma$ and $\bar B\to X_{c,u}\,l\,\bar\nu_l$. The ``$|V_{ub}|$ vs.\ $\sin2\beta$ puzzle", i.e., the fact that there used to be a tension between the determination of $|V_{ub}|$ from inclusive semileptonic decays as compared with the indirect value inferred from the measurement of the angle $\beta$ of the unitarity triangle, is resolved when only model-independent information is used in the extraction of the $b$-quark mass and other heavy-quark parameters. The new value of $|V_{ub}|$ obtained in this way is in good agreement with the determination from exclusive decays.

\subsection*{Acknowledgments}

I am grateful to the organizers of the 2007 Lepton-Photon Symposium for the invitation to deliver this talk and for the flawless organization of this important conference. Many thanks go to Elisabetta Barberio and Francesca Di Lodovico, whose help was instrumental in getting the most up-to-date averages for $m_b$ and $|V_{ub}|$.


\begin{thebibliography}{199}

\bibitem{Buchalla:1995vs}
For a review, see:
  G.~Buchalla, A.~J.~Buras and M.~E.~Lautenbacher,
  Rev.\ Mod.\ Phys.\  {\bf 68}, 1125 (1996)
  [arXiv:hep-ph/9512380].

\bibitem{Isgur:1989ed}
  N.~Isgur and M.~B.~Wise,
  Phys.\ Lett.\  B {\bf 237}, 527 (1990).

\bibitem{Neubert:1993mb}
For a review, see:
  M.~Neubert,
  Phys.\ Rept.\  {\bf 245}, 259 (1994)
  [arXiv:hep-ph/9306320].

\bibitem{Bigi:1992su}
  I.~I.~Y.~Bigi, N.~G.~Uraltsev and A.~I.~Vainshtein,
  Phys.\ Lett.\  B {\bf 293}, 430 (1992)
  [Erratum-ibid.\  B {\bf 297}, 477 (1993)]
  [arXiv:hep-ph/9207214].

\bibitem{Bigi:1993fe}
  I.~I.~Y.~Bigi, M.~A.~Shifman, N.~G.~Uraltsev and A.~I.~Vainshtein,
  Phys.\ Rev.\ Lett.\  {\bf 71}, 496 (1993)
  [arXiv:hep-ph/9304225].

\bibitem{Manohar:1993qn}
  A.~V.~Manohar and M.~B.~Wise,
  Phys.\ Rev.\  D {\bf 49}, 1310 (1994)
  [arXiv:hep-ph/9308246].

\bibitem{Sterman:1986aj}
  G.~Sterman,
  Nucl.\ Phys.\  B {\bf 281}, 310 (1987).

\bibitem{Catani:1989ne}
  S.~Catani and L.~Trentadue,
  Nucl.\ Phys.\  B {\bf 327}, 323 (1989).

\bibitem{Korchemsky:1992xv}
  G.~P.~Korchemsky and G.~Marchesini,
  Nucl.\ Phys.\  B {\bf 406}, 225 (1993)
  [arXiv:hep-ph/9210281].

\bibitem{Bauer:2000yr}
  C.~W.~Bauer, S.~Fleming, D.~Pirjol and I.~W.~Stewart,
  Phys.\ Rev.\  D {\bf 63}, 114020 (2001)
  [arXiv:hep-ph/0011336].

\bibitem{Bauer:2001yt}
  C.~W.~Bauer, D.~Pirjol and I.~W.~Stewart,
  Phys.\ Rev.\  D {\bf 65}, 054022 (2002)
  [arXiv:hep-ph/0109045].

\bibitem{Beneke:2002ph}
  M.~Beneke, A.~P.~Chapovsky, M.~Diehl and T.~Feldmann,
  Nucl.\ Phys.\  B {\bf 643}, 431 (2002)
  [arXiv:hep-ph/0206152].

\bibitem{Hill:2002vw}
  R.~J.~Hill and M.~Neubert,
  Nucl.\ Phys.\  B {\bf 657}, 229 (2003)
  [arXiv:hep-ph/0211018].

\bibitem{Manohar:2003vb}
  A.~V.~Manohar,
  Phys.\ Rev.\  D {\bf 68}, 114019 (2003)
  [arXiv:hep-ph/0309176].

\bibitem{Becher:2006mr}
  T.~Becher, M.~Neubert and B.~D.~Pecjak,
  JHEP {\bf 0701}, 076 (2007)
  [arXiv:hep-ph/0607228].

\bibitem{Becher:2007ty}
  T.~Becher, M.~Neubert and G.~Xu,
  arXiv:0710.0680 [hep-ph].

\bibitem{Korchemsky:1999kt}
  G.~P.~Korchemsky and G.~Sterman,
  Nucl.\ Phys.\  B {\bf 555}, 335 (1999)
  [arXiv:hep-ph/9902341].

\bibitem{Neubert:1993ch}
  M.~Neubert,
  Phys.\ Rev.\  D {\bf 49}, 3392 (1994)
  [arXiv:hep-ph/9311325].

\bibitem{Neubert:1993um}
  M.~Neubert,
  Phys.\ Rev.\  D {\bf 49}, 4623 (1994)
  [arXiv:hep-ph/9312311].

\bibitem{Bigi:1993ex}
  I.~I.~Y.~Bigi, M.~A.~Shifman, N.~G.~Uraltsev and A.~I.~Vainshtein,
  Int.\ J.\ Mod.\ Phys.\  A {\bf 9}, 2467 (1994)
  [arXiv:hep-ph/9312359].

\bibitem{Korchemsky:1994jb}
  G.~P.~Korchemsky and G.~Sterman,
  Phys.\ Lett.\  B {\bf 340}, 96 (1994)
  [arXiv:hep-ph/9407344].

\bibitem{Bosch:2004th}
  S.~W.~Bosch, B.~O.~Lange, M.~Neubert and G.~Paz,
  Nucl.\ Phys.\  B {\bf 699}, 335 (2004)
  [arXiv:hep-ph/0402094].

\bibitem{Beneke:2000wa}
  M.~Beneke and T.~Feldmann,
  Nucl.\ Phys.\  B {\bf 592}, 3 (2001)
  [arXiv:hep-ph/0008255].

\bibitem{Ball:2004ye}
  P.~Ball and R.~Zwicky,
  Phys.\ Rev.\  D {\bf 71}, 014015 (2005)
  [arXiv:hep-ph/0406232].

\bibitem{Ball:2004rg}
  P.~Ball and R.~Zwicky,
  Phys.\ Rev.\  D {\bf 71}, 014029 (2005)
  [arXiv:hep-ph/0412079].

\bibitem{Beneke:1999br}
  M.~Beneke, G.~Buchalla, M.~Neubert and C.~T.~Sachrajda,
  Phys.\ Rev.\ Lett.\  {\bf 83}, 1914 (1999)
  [arXiv:hep-ph/9905312].

\bibitem{Beneke:2000ry}
  M.~Beneke, G.~Buchalla, M.~Neubert and C.~T.~Sachrajda,
  Nucl.\ Phys.\  B {\bf 591}, 313 (2000)
  [arXiv:hep-ph/0006124].

\bibitem{Beneke:2001ev}
  M.~Beneke, G.~Buchalla, M.~Neubert and C.~T.~Sachrajda,
  Nucl.\ Phys.\  B {\bf 606}, 245 (2001)
  [arXiv:hep-ph/0104110].

\bibitem{Beneke:2002jn}
  M.~Beneke and M.~Neubert,
  Nucl.\ Phys.\  B {\bf 651}, 225 (2003)
  [arXiv:hep-ph/0210085].

\bibitem{Beneke:2003zv}
  M.~Beneke and M.~Neubert,
  Nucl.\ Phys.\  B {\bf 675}, 333 (2003)
  [arXiv:hep-ph/0308039].

\bibitem{Beneke:2001at}
  M.~Beneke, T.~Feldmann and D.~Seidel,
  Nucl.\ Phys.\  B {\bf 612}, 25 (2001)
  [arXiv:hep-ph/0106067].

\bibitem{Bosch:2001gv}
  S.~W.~Bosch and G.~Buchalla,
  Nucl.\ Phys.\  B {\bf 621}, 459 (2002)
  [arXiv:hep-ph/0106081].

\bibitem{Beneke:2005vv}
  M.~Beneke and S.~J\"ager,
  Nucl.\ Phys.\  B {\bf 751}, 160 (2006)
  [arXiv:hep-ph/0512351].

\bibitem{Beneke:2006mk}
  M.~Beneke and S.~J\"ager,
  Nucl.\ Phys.\  B {\bf 768}, 51 (2007)
  [arXiv:hep-ph/0610322].

\bibitem{Bell:2007tv}
  G.~Bell,
  arXiv:0705.3127 [hep-ph].

\bibitem{Becher:2005fg}
  T.~Becher, R.~J.~Hill and M.~Neubert,
  Phys.\ Rev.\  D {\bf 72}, 094017 (2005)
  [arXiv:hep-ph/0503263].

\bibitem{Ali:2007sj}
  A.~Ali, B.~D.~Pecjak and C.~Greub,
  arXiv:0709.4422 [hep-ph].

\bibitem{Lee:2004ja}
  K.~S.~M.~Lee and I.~W.~Stewart,
  Nucl.\ Phys.\  B {\bf 721}, 325 (2005)
  [arXiv:hep-ph/0409045].

\bibitem{Bosch:2004cb}
  S.~W.~Bosch, M.~Neubert and G.~Paz,
  JHEP {\bf 0411}, 073 (2004)
  [arXiv:hep-ph/0409115].

\bibitem{Beneke:2004in}
  M.~Beneke, F.~Campanario, T.~Mannel and B.~D.~Pecjak,
  JHEP {\bf 0506}, 071 (2005)
  [arXiv:hep-ph/0411395].

\bibitem{Szczepaniak:1990dt}
  A.~Szczepaniak, E.~M.~Henley and S.~J.~Brodsky,
  Phys.\ Lett.\  B {\bf 243}, 287 (1990).

\bibitem{Burdman:1992hg}
  G.~Burdman and J.~F.~Donoghue,
  Phys.\ Lett.\  B {\bf 270}, 55 (1991).

\bibitem{Bauer:2002aj}
  C.~W.~Bauer, D.~Pirjol and I.~W.~Stewart,
  Phys.\ Rev.\  D {\bf 67}, 071502 (2003)
  [arXiv:hep-ph/0211069].

\bibitem{Beneke:2003pa}
  M.~Beneke and T.~Feldmann,
  Nucl.\ Phys.\  B {\bf 685}, 249 (2004)
  [arXiv:hep-ph/0311335].

\bibitem{Lange:2003pk}
  B.~O.~Lange and M.~Neubert,
  Nucl.\ Phys.\  B {\bf 690}, 249 (2004)
  [Erratum-ibid.\  B {\bf 723}, 201 (2005)]
  [arXiv:hep-ph/0311345].

\bibitem{Manohar:2006nz}
  A.~V.~Manohar and I.~W.~Stewart,
  Phys.\ Rev.\  D {\bf 76}, 074002 (2007)
  [arXiv:hep-ph/0605001].

\bibitem{Arnesen:2006vb}
  C.~M.~Arnesen, Z.~Ligeti, I.~Z.~Rothstein and I.~W.~Stewart,
  arXiv:hep-ph/0607001.

\bibitem{Jain:2007dy}
  A.~Jain, I.~Z.~Rothstein and I.~W.~Stewart,
  arXiv:0706.3399 [hep-ph].

\bibitem{Bauer:2004tj}
  C.~W.~Bauer, D.~Pirjol, I.~Z.~Rothstein and I.~W.~Stewart,
  Phys.\ Rev.\  D {\bf 70}, 054015 (2004)
  [arXiv:hep-ph/0401188].

\bibitem{Beneke:2004bn}
  M.~Beneke, G.~Buchalla, M.~Neubert and C.~T.~Sachrajda,
  Phys.\ Rev.\  D {\bf 72}, 098501 (2005)
  [arXiv:hep-ph/0411171].

\bibitem{Misiak:2006zs}
  M.~Misiak {\it et al.},
  Phys.\ Rev.\ Lett.\  {\bf 98}, 022002 (2007)
  [arXiv:hep-ph/0609232].

\bibitem{Becher:2005pd}
  T.~Becher and M.~Neubert,
  Phys.\ Lett.\  B {\bf 633}, 739 (2006)
  [arXiv:hep-ph/0512208].

\bibitem{Becher:2006qw}
  T.~Becher and M.~Neubert,
  Phys.\ Lett.\  B {\bf 637}, 251 (2006)
  [arXiv:hep-ph/0603140].

\bibitem{Becher:2006pu}
  T.~Becher and M.~Neubert,
  Phys.\ Rev.\ Lett.\  {\bf 98}, 022003 (2007)
  [arXiv:hep-ph/0610067].

\bibitem{Bobeth:1999mk}
  C.~Bobeth, M.~Misiak and J.~Urban,
  Nucl.\ Phys.\  B {\bf 574}, 291 (2000)
  [arXiv:hep-ph/9910220].

\bibitem{Misiak:2004ew}
  M.~Misiak and M.~Steinhauser,
  Nucl.\ Phys.\  B {\bf 683}, 277 (2004)
  [arXiv:hep-ph/0401041].

\bibitem{Bobeth:1999ww}
  C.~Bobeth, M.~Misiak and J.~Urban,
  Nucl.\ Phys.\  B {\bf 567}, 153 (2000)
  [arXiv:hep-ph/9904413].

\bibitem{Gorbahn:2004my}
  M.~Gorbahn and U.~Haisch,
  Nucl.\ Phys.\  B {\bf 713}, 291 (2005)
  [arXiv:hep-ph/0411071].

\bibitem{Gorbahn:2005sa}
  M.~Gorbahn, U.~Haisch and M.~Misiak,
  Phys.\ Rev.\ Lett.\  {\bf 95}, 102004 (2005)
  [arXiv:hep-ph/0504194].

\bibitem{Czakon:2006ss}
  M.~Czakon, U.~Haisch and M.~Misiak,
  JHEP {\bf 0703}, 008 (2007)
  [arXiv:hep-ph/0612329].

\bibitem{Melnikov:2005bx}
  K.~Melnikov and A.~Mitov,
  Phys.\ Lett.\  B {\bf 620}, 69 (2005)
  [arXiv:hep-ph/0505097].

\bibitem{Blokland:2005uk}
  I.~Blokland, A.~Czarnecki, M.~Misiak, M.~Slusarczyk and F.~Tkachov,
  Phys.\ Rev.\  D {\bf 72}, 033014 (2005)
  [arXiv:hep-ph/0506055].

\bibitem{Asatrian:2006ph}
  H.~M.~Asatrian, A.~Hovhannisyan, V.~Poghosyan, T.~Ewerth, C.~Greub and T.~Hurth,
  Nucl.\ Phys.\  B {\bf 749}, 325 (2006)
  [arXiv:hep-ph/0605009].

\bibitem{Asatrian:2006sm}
  H.~M.~Asatrian, T.~Ewerth, A.~Ferroglia, P.~Gambino and C.~Greub,
  Nucl.\ Phys.\  B {\bf 762}, 212 (2007)
  [arXiv:hep-ph/0607316].

\bibitem{Misiak:2006ab}
  M.~Misiak and M.~Steinhauser,
  Nucl.\ Phys.\  B {\bf 764}, 62 (2007)
  [arXiv:hep-ph/0609241].

\bibitem{Ligeti:1999ea}
  Z.~Ligeti, M.~E.~Luke, A.~V.~Manohar and M.~B.~Wise,
  Phys.\ Rev.\  D {\bf 60}, 034019 (1999)
  [arXiv:hep-ph/9903305].

\bibitem{Bieri:2003ue}
  K.~Bieri, C.~Greub and M.~Steinhauser,
  Phys.\ Rev.\  D {\bf 67}, 114019 (2003)
  [arXiv:hep-ph/0302051].

\bibitem{Lee:2006wn}
  S.~J.~Lee, M.~Neubert and G.~Paz,
  Phys.\ Rev.\  D {\bf 75}, 114005 (2007)
  [arXiv:hep-ph/0609224].

\bibitem{inprep}
S.~J.~Lee, M.~Neubert and G.~Paz, in preparation.

\bibitem{Neubert:2004dd}
  M.~Neubert,
  Eur.\ Phys.\ J.\  C {\bf 40}, 165 (2005)
  [arXiv:hep-ph/0408179].

\bibitem{Neubert:2005nt}
  M.~Neubert,
  Phys.\ Rev.\  D {\bf 72}, 074025 (2005)
  [arXiv:hep-ph/0506245].

\bibitem{Buchmuller:2005zv}
  O.~Buchm\"uller and H.~Fl\"acher,
  Phys.\ Rev.\  D {\bf 73}, 073008 (2006)
  [arXiv:hep-ph/0507253].

\bibitem{Andersen:2006hr}
  J.~R.~Andersen and E.~Gardi,
  JHEP {\bf 0701}, 029 (2007)
  [arXiv:hep-ph/0609250].

\bibitem{Barberio:2007cr}
  E.~Barberio {\it et al.}  [Heavy Flavor Averaging Group (HFAG)
                  Collaboration],
  arXiv:0704.3575 [hep-ex];
for LP07 updates, see: www.slac.stanford.edu/xorg/hfag.

\bibitem{Bauer:2002sh}
  C.~W.~Bauer, Z.~Ligeti, M.~Luke and A.~V.~Manohar,
  Phys.\ Rev.\  D {\bf 67}, 054012 (2003)
  [arXiv:hep-ph/0210027].

\bibitem{Bigi:2003zg}
  I.~I.~Bigi and N.~Uraltsev,
  Phys.\ Lett.\  B {\bf 579}, 340 (2004)
  [arXiv:hep-ph/0308165].

\bibitem{Gambino:2004qm}
  P.~Gambino and N.~Uraltsev,
  Eur.\ Phys.\ J.\  C {\bf 34}, 181 (2004)
  [arXiv:hep-ph/0401063].

\bibitem{Bauer:2004ve}
  C.~W.~Bauer, Z.~Ligeti, M.~Luke, A.~V.~Manohar and M.~Trott,
  Phys.\ Rev.\  D {\bf 70}, 094017 (2004)
  [arXiv:hep-ph/0408002].

\bibitem{Anastasiou:2005pn}
  C.~Anastasiou, K.~Melnikov and F.~Petriello,
  JHEP {\bf 0709}, 014 (2007)
  [arXiv:hep-ph/0505069].

\bibitem{Becher:2007tk}
  T.~Becher, H.~Boos and E.~Lunghi,
  arXiv:0708.0855 [hep-ph].

\bibitem{Benson:2003kp}
  D.~Benson, I.~I.~Bigi, T.~Mannel and N.~Uraltsev,
  Nucl.\ Phys.\  B {\bf 665}, 367 (2003)
  [arXiv:hep-ph/0302262].

\bibitem{Hoang:1998hm}
  A.~H.~Hoang, Z.~Ligeti and A.~V.~Manohar,
  Phys.\ Rev.\  D {\bf 59}, 074017 (1999)
  [arXiv:hep-ph/9811239].

\bibitem{Neubert:2004sp}
  M.~Neubert,
  Phys.\ Lett.\  B {\bf 612}, 13 (2005)
  [arXiv:hep-ph/0412241].

\bibitem{Kuhn:2007vp}
  J.~H.~K\"uhn, M.~Steinhauser and C.~Sturm,
  Nucl.\ Phys.\  B {\bf 778}, 192 (2007)
  [arXiv:hep-ph/0702103].

\bibitem{Kagan:1998ym}
  A.~L.~Kagan and M.~Neubert,
  Eur.\ Phys.\ J.\  C {\bf 7}, 5 (1999)
  [arXiv:hep-ph/9805303].

\bibitem{Benson:2004sg}
  D.~Benson, I.~I.~Bigi and N.~Uraltsev,
  Nucl.\ Phys.\  B {\bf 710}, 371 (2005)
  [arXiv:hep-ph/0410080].

\bibitem{Bosch:2004bt}
  S.~W.~Bosch, B.~O.~Lange, M.~Neubert and G.~Paz,
  Phys.\ Rev.\ Lett.\  {\bf 93}, 221801 (2004)
  [arXiv:hep-ph/0403223].

\bibitem{Lange:2005yw}
  B.~O.~Lange, M.~Neubert and G.~Paz,
  Phys.\ Rev.\  D {\bf 72}, 073006 (2005)
  [arXiv:hep-ph/0504071].

\bibitem{Leibovich:1999xf}
  A.~K.~Leibovich, I.~Low and I.~Z.~Rothstein,
  Phys.\ Rev.\  D {\bf 61}, 053006 (2000)
  [arXiv:hep-ph/9909404].

\bibitem{Leibovich:2000ey}
  A.~K.~Leibovich, I.~Low and I.~Z.~Rothstein,
  Phys.\ Lett.\  B {\bf 486}, 86 (2000)
  [arXiv:hep-ph/0005124].

\bibitem{Lange:2005qn}
  B.~O.~Lange, M.~Neubert and G.~Paz,
  JHEP {\bf 0510}, 084 (2005)
  [arXiv:hep-ph/0508178].

\bibitem{Lange:2005xz}
  B.~O.~Lange,
  JHEP {\bf 0601}, 104 (2006)
  [arXiv:hep-ph/0511098].

\bibitem{inprep2}
H.~M.~Asatrian, C.~Greub, M.~Neubert and B.~D.~Pecjak, in preparation.

\bibitem{Gardi:2004ia}
  E.~Gardi,
  JHEP {\bf 0404}, 049 (2004)
  [arXiv:hep-ph/0403249].

\bibitem{Andersen:2005bj}
  J.~R.~Andersen and E.~Gardi,
  JHEP {\bf 0506}, 030 (2005)
  [arXiv:hep-ph/0502159].

\bibitem{Andersen:2005mj}
  J.~R.~Andersen and E.~Gardi,
  JHEP {\bf 0601}, 097 (2006)
  [arXiv:hep-ph/0509360].

\bibitem{Gambino:2007rp}
  P.~Gambino, P.~Giordano, G.~Ossola and N.~Uraltsev,
  JHEP {\bf 0710}, 058 (2007)
  [arXiv:0707.2493 [hep-ph]].

\bibitem{Gambino:2006wk}
  P.~Gambino, E.~Gardi and G.~Ridolfi,
  JHEP {\bf 0612}, 036 (2006)
  [arXiv:hep-ph/0610140].

\bibitem{Okamoto:2004xg}
  M.~Okamoto {\it et al.},
  Nucl.\ Phys.\ Proc.\ Suppl.\  {\bf 140}, 461 (2005)
  [arXiv:hep-lat/0409116].

\bibitem{Dalgic:2006dt}
  E.~Dalgic, A.~Gray, M.~Wingate, C.~T.~H.~Davies, G.~P.~Lepage and J.~Shigemitsu,
  Phys.\ Rev.\  D {\bf 73}, 074502 (2006)
  [Erratum-ibid.\  D {\bf 75}, 119906 (2007)]
  [arXiv:hep-lat/0601021].

\end{thebibliography}
\end{document}